\documentclass[
superscriptaddress,
nofootinbib, twocolumn,
amsmath,amssymb,aps, preprintnumbers,
floatfix, eprint ]{revtex4-2}

\usepackage[T1]{fontenc} 
\usepackage{graphicx}
\usepackage{tikz}
\usepackage[colorlinks=true,linktocpage=true,linkcolor=blue,citecolor=blue]{hyperref}
\usepackage{cleveref}

\def\q{{\boldsymbol q}}
\def\k{{\boldsymbol k}}

\def\x{{\boldsymbol x}}
\def\p{{\boldsymbol p}}

\newcommand{\rmd}{{\rm d}}

\begin{document}
\title{An Equilibrating Parton Shower for Jet Quenching and Medium Response}

\author{Ismail Soudi}
\email{isma@physik.uni-bielefeld.de}
\affiliation{Fakult\"at f\"ur Physik, Universit\"at Bielefeld, D-33615 Bielefeld, Germany}

\author{Adam Takacs}
\email{adam.takacs@cern.ch}
\affiliation{CERN, Theoretical Physics Department, CH-1211 Geneva 23, Switzerland}
\affiliation{Department of Physics and Technology, University of Bergen, PO-7803, 5020 Bergen, Norway}

\preprint{CERN-TH-2026-180}

\begin{abstract}
    In the weak-coupling approach, thermalization of the quark-gluon plasma and the evolution of jet-like perturbations in the medium can be described by the same underlying QCD effective kinetic theory (EKT).
    In this work, we show how the EKT reduces to the standard energy-loss and angular-broadening evolution of jet-quenching in the high-energy regime, while providing a consistent description near the thermal scales.
    Building on this, we derive a parton shower directly from EKT that reproduces the full linearized Boltzmann equation, combining elastic and inelastic collisions to capture both hard-parton energy loss and thermal equilibration on an equal footing.
    This shower lets us extend EKT to inhomogeneous perturbations, connecting jet evolution to the medium's hydrodynamic response, including wakes, Mach cones, and multi-particle correlations. 
\end{abstract}
\maketitle

\textbf{Introduction} --- 
Among the clearest signatures of quark–gluon plasma (QGP) produced in ultrarelativistic heavy-ion collisions is jet quenching~\cite{Busza:2018rrf,Harris:2023tti}, the modification of jets relative to their proton–proton counterparts~\cite{Armesto:2015ioy,Connors:2017ptx,Cunqueiro:2021wls}, which allows us to infer QGP properties~\cite{JET:2013cls,JETSCAPE:2021ehl,JETSCAPE:2024cqe}. At weak coupling, this quenching traces to medium-modified scattering amplitudes and their associated renormalization-group-like evolution equations~\cite{Casalderrey-Solana:2007knd,Majumder:2010qh,Mehtar-Tani:2013pia,Qin:2015srf,Mehtar-Tani:2025rty}, encoded in jet quenching Monte Carlo (MC) generators~\cite{Majumder:2013re,Zapp:2013vla,Caucal:2018dla} that capture an extensive set of hadron and jet observables~\cite{CMS:2021vui,ATLAS:2022vii,ATLAS:2023iad,ALICE:2023waz,ALICE:2024fip,CMS:2026bmz}.

Connecting jet evolution with the heavy-ion bulk remains a current frontier~\cite{Schenke:2010rr,Nijs:2020ors,Wu:2021fjf}, demanding a self-consistent accounting of energy-momentum deposition and medium response~\cite{Cao:2020wlm,He:2022evt,Barreto:2022ulg,Cao:2024pxc,vanderSchee:2025hoe}. Such models commonly adopt a switching scale of the order of 1 GeV, treating partons below this momentum as instantaneously thermalized sources of hydrodynamic wakes~\cite{Cao:2020wlm,Cao:2024pxc}. However, whether equilibration is achieved in fact remains an open question. Strongly coupled descriptions point toward rapid equilibration~\cite{Gubser:2007ga,Chesler:2007sv,Casalderrey-Solana:2020rsj,Pablos:2022piv} and enjoy considerable success, but rest on phenomenological input rather than a first-principles foundation. In recent years, the search for a jet-induced medium response inspired many studies~\cite{Pablos:2019ngg,Yang:2022nei,Yang:2023dwc,Bossi:2024qho,Yang:2025dqu} and measurements~\cite{CMS:2021otx,ATLAS:2024prm,CMS:2025dua,CMS:2026mur} encouraging a more rigorous formulation.

High-temperature QCD effective kinetic theory (AMY-EKT)~\cite{Arnold:2002zm} provides microscopic understanding of how heavy-ion collisions equilibrate in line with the ``bottom-up'' picture~\cite{Baier:2000sb,Schlichting:2019abc,Berges:2020fwq}.
Early steps toward describing jet-like perturbations have already been taken~\cite{Schlichting:2020lef,Mehtar-Tani:2022zwf,Zhou:2024ysb,Boguslavski:2025ylx} by solving the EKT numerically.
These analyses exposed pronounced departures from equilibrium precisely in the relevant $1-5$~GeV region identified above.
This opens a route to studying out-of-equilibrium jet evolution, especially pertinent to the recent light-ion program at LHC and RHIC~\cite{Citron:2018lsq,Brewer:2021kiv,CMS:2025bta,ALICE:2026zck,ATLAS:2026fos,STAR:2026nfy,sPH-CONF-JET-2026-02}, where less thermalization is expected compared to PbPb and AuAu collisions.

The same EKT framework describes both the thermalization of the medium and the evolution of jet-like perturbations.
In the high-energy limit ($p\gg T$) these evolution equations reduce to DGLAP-like evolution~\cite{Mehtar-Tani:2018zba}. 
However, as the jet fragments approach the thermal scale, other processes become important, and the EKT provides a completion of the evolution equations in this regime.
Establishing this connection points towards a single, coherent framework for studying jet equilibration and medium response.

However, realizing this connection in practice requires a parton shower that faithfully reproduces the EKT. Earlier efforts to recast the linearized Boltzmann equation in MC form~\cite{Zapp:2008gi,Auvinen:2009qm,Schenke:2009gb,He:2015pra,He:2018xjv,DEramo:2018eoy} were hampered by several shortcomings that preclude detailed balance and thus equilibration, e.g., secondary collisions are usually left out, and $2\!\to\!1$ merging is omitted. We derive a new parton shower, starting from the EKT, that reproduces the full linearized Boltzmann equation, including the complete set of $1\!\leftrightarrow\!2$ and $2\!\leftrightarrow\!2$ collisions with quantum-statistical factors. This yields a consistent description of both the energy-loss evolution and the thermalization of the soft fragments, capturing the full medium response.

Going beyond previous EKT studies, we further extend our framework to inhomogeneous perturbations in three spatial dimensions, establishing a direct connection to the subsequent hydrodynamic evolution of density and momentum excitations in the medium, including hydrodynamic wakes and Mach cones.

Even though our focus is on thermalization in QCD and heavy-ion collisions, the proposed framework applies broadly to nonequilibrium problems governed by the Boltzmann equation, such as reheating in the early universe and neutrino transport in core-collapse supernovae, where solving the fully inhomogeneous kinetic equation is likewise the bottleneck~\cite{Kolb:1990vq,Konstandin:2013caa,Mezzacappa:2020oyq}.

\textbf{The EKT framework}~\cite{Arnold:2002zm} --- 
The evolution of a small perturbation $\delta f(t,\x,\p)=(2\pi)^3\rmd^6 N/(\rmd^3\x\rmd^3\p)$ on top of an equilibrium background $n(p)$ is described by the linearized Boltzmann equation~\cite{Schlichting:2020lef,Zhou:2024ysb},
\begin{equation}\label{eq:linBoltzmann}
    \left(\partial_t+\hat{\p}\cdot
    \nabla_{\x}\right)\delta f_a(t,\x,\p)
    = \delta C_a^{2\leftrightarrow2} + \delta C_a^{1\leftrightarrow2}\,,
\end{equation}
where $\hat{\p}=\p/p$, $a$ labels quarks and gluons, and $\delta C^{i\leftrightarrow j}_a[\{n,\delta f\}](t,\x,\p)$ is the linearized collision describing elastic $2\leftrightarrow2$ and inelastic $1\leftrightarrow2$ processes. The AMY collisions are summarized in App.~\ref{sec:Collision_kernels}; however, our derivation holds for any transport equation of this form.

Ref.~\cite{Soudi:2025lei} established this shower construction and its equilibration for purely inelastic ($1\leftrightarrow2$) processes; elastic scattering is needed both for the transverse-momentum broadening central to jet quenching and for genuine isotropization, so here we retain the full $2\leftrightarrow2$ collisions alongside $1\leftrightarrow2$. Following~\cite{Soudi:2025lei}, we separate the collision terms into momentum-changing (real) and momentum-preserving (virtual) contributions,
\begin{equation}
    \begin{split}
        \delta C^{1\leftrightarrow2}_a & = \delta C^{1\leftrightarrow2,{\rm r}}_a[\delta f] - \delta C^{1\leftrightarrow2,{\rm v}}_a[n]\cdot\delta f_a(t,\p)\,, \\
        \delta C^{2\leftrightarrow2}_a & = \delta C^{2\leftrightarrow2,{\rm r}}_a[\delta f] - \delta C^{2\leftrightarrow2,{\rm v}}_a[n]\cdot\delta f_a(t,\p)\,.
    \end{split}
\end{equation}
This allows us to rewrite the spatially homogeneous part of \cref{eq:linBoltzmann} (or, equivalently, its spatial integral) as
\begin{equation}\label{eq:PartonShower}
    \begin{split}
        \delta f_a(t,\p) & = \Delta_a(t-t_0,\p)\delta f_{a0} + \int_{t_0}^t\rmd t'\Delta_a(t-t',\p)                                           \\
                         & \times(\delta C^{1\leftrightarrow2,{\rm r}}_a[\delta f] + \delta C^{2\leftrightarrow2,{\rm r}}_a[\delta f])\,,
    \end{split}
\end{equation}
where $\delta f_{a0}=\delta f_a(t_0,\p)$ is the initial condition and
\begin{equation}\label{eq:Sudakov}
    \Delta_a(t-t_0,\p) = \exp\left[-\int^t_{t_0}\rmd t' (\delta C^{1\leftrightarrow2,{\rm v}}_a + \delta C^{2\leftrightarrow2,{\rm v}}_a)\right]
\end{equation}
is the no-collision (Sudakov) factor. The first term in \cref{eq:PartonShower} describes the uninterrupted propagation of the initial condition, while the second term describes the uninterrupted evolution following the last collision at time $t'$, which involved $\delta f_a(t',\p)$. \Cref{eq:PartonShower} can therefore be iterated by substituting its own right-hand side for $\delta f_a$ inside the collision integrals, systematically generating an arbitrary number of collisions. Such algorithms are well known in pQCD, referred to as parton showers~\cite{Lonnblad:2012hz,Hoche:2014rga,Bierlich:2022pfr,Bewick:2023tfi,Sherpa:2024mfk}.

Finally, \cref{eq:PartonShower} can be extended to inhomogeneous perturbations $\delta f(t,\x,\p)$, as shown in App.~\ref{sec:Inhomogeneous_perturbations}. The idea is to rewrite \cref{eq:linBoltzmann} in comoving coordinates, in which the streaming term $\hat{\p}\cdot\nabla_{\x}$ is absent and the solution is given by \cref{eq:PartonShower}, except that the collision terms acquire a non-trivial spatial dependence. Transforming back to lab-frame coordinates, we obtain
\begin{align} \label{eq:PartonShower_inhomo}
     & \delta f(t,\x,\p) = \Delta(t-t_0,\p)\, \delta f(t_0,\x-\hat{\p}(t-t_0),\p)\\
     & \quad +\int^t_{t_0}\rmd t'\,\Delta(t-t',\p)\,\delta C^{\rm r}\big[\{n,\delta f(t',\x-\hat{\p}(t-t'),\k)\}\big]\,, \nonumber
\end{align}
where indices are omitted for simplicity.
The first line describes the uninterrupted, free propagation of the initial condition, while the second line describes particles arriving at $(t,\x)$ for which a collision must have occurred at $(t',\x-\hat{\p}(t-t'))$, where the particle previously had momentum $\k$.
\Cref{eq:PartonShower_inhomo} yields a simple and intuitive picture: particles propagate freely between collisions, at which their momentum and direction change. Although this picture is widely used in MC generators, to our knowledge, its explicit realization has not been derived in the EKT. Remarkably, \cref{eq:PartonShower_inhomo} is equivalent to \cref{eq:PartonShower}, and thus retains detailed balance, equilibration, quantum statistics, and energy-momentum conservation.

\textbf{High-energy limit of the EKT and its connection to jet quenching} --- 
It is known that \cref{eq:linBoltzmann} drives perturbations toward thermal equilibrium. We show that it simultaneously reproduces the evolution equations found in the BDMPS-Z formalism to describe the broadening and radiation of a hard parton. In simple terms, \cref{eq:linBoltzmann} bridges jet quenching and thermalization: deviations from standard parton evolution are suppressed in the $p\gg T$ regime, with EKT modifying the dynamics only in the region $p\sim T$.

In the $p\gg T$ limit, \cref{eq:linBoltzmann} simplifies drastically~\cite{Soudi:2025lei}
\begin{equation}
    \begin{split}
        \delta C^{1\leftrightarrow2}(\p) & \approx \int_0^1\rmd z \Big[\tfrac{1}{z^3}\Gamma(z,\tfrac{\p}{z})\delta f_{p/z} - \tfrac12\Gamma(z,\p)\delta f_p \Big]\,,
    \end{split}
    \label{eq:dC12_highenergy}
\end{equation}
where $\Gamma$ is the high-energy limit of the AMY splitting rate, equal to the medium-induced splitting rate of BDMPS-Z~\cite{Baier:1996kr,Zakharov:1996fv,Arnold:2008iy}.
The elastic collision term in the same limit is
\begin{equation}
    \delta C^{2\leftrightarrow2}(\p) \approx \int_{\k,\q}W\cdot\left[\delta f_{p+q}n_{k-q}\bar n_k - \delta f_p \bar n_{k-q} n_k\right]\,,
\end{equation}
where $\q=\p'-\p$, and
\begin{equation}
    W(\p,\k;\q) \equiv
    \frac{2\pi\,\delta(|\p|{+}|\k|{-}|\p{+}\q|{-}|\k{-}\q|)\,
        |\mathcal{M}|^2}
    {2\nu_a\,16\,|\p|\,|\p{+}\q|\,|\k|\,|\k{-}\q|}\,.
    \label{eq:W}
\end{equation}

In the soft-scattering limit $q\ll k$, the energy delta reduces to $\delta((\hat{\p}-\hat{\k})\cdot\q)$, implying that both energy and longitudinal momentum exchange are negligible compared to $T$, so that $\delta f(\p+\q)\approx\delta f(\p+\q_\perp)$. This leads to
\begin{equation}
    \delta C_a^{2\leftrightarrow2}(\p)\approx\int_{\q_\perp}\gamma_a(q_
    \perp)\Big[\delta f_a(t,\p+\q_\perp)-\delta f_a(t,\p)\Big]\,,
    \label{eq:Broadening_highenergy}
\end{equation}
where $\int_{\q_\perp}\equiv\int\frac{\rmd^2\q_\perp}{(2\pi)^2}$, and the broadening kernel, evaluated in the hard-thermal-loop (HTL) approximation~\cite{Arnold:2008vd},
\begin{equation}
    \begin{split}
        \gamma_a(q_\perp) & \approx
        \begin{cases}
            \tfrac{g^2C_am_D^2T}{q_\perp^2(q_\perp^2+m_D^2)}\,,                     & q_\perp\ll gT\,, \\
            \tfrac{g^4T^3C_a}{q_\perp^4}\tfrac{\zeta(3)}{\zeta(2)}(1+\tfrac{n_f}{4})\,, & q_\perp\gg gT\,,
        \end{cases}
    \end{split}
\end{equation}
where $m_D^2=g^2T^2(1+n_f/6)$, $C_g=3$, $C_q=4/3$. \Cref{eq:Broadening_highenergy} can be further simplified into a Gaussian broadening (diffusion) by expanding $\delta f(\p+\q_\perp)$ in $\q_\perp$; alternatively, one may retain the Coulomb tail $\gamma\sim q_\perp^{-4}$ characteristic of rare, wide-angle scatterings.

The linearized Boltzmann equation, with the high-energy collision terms \cref{eq:dC12_highenergy} and \cref{eq:Broadening_highenergy}, reproduces the evolution equations for parton energy loss and broadening in the BDMPS-Z formalism~\cite{Blaizot:2013vha,Blaizot:2014ula,Blaizot:2015jea}, widely used in jet quenching phenomenology~\cite{Mehtar-Tani:2016aco,Zakharov:2020whb,Caucal:2021cfb,Pablos:2022mrx,Cunqueiro:2022svx,Apolinario:2026hff,Duan:2026nvr} and in MC implementations~\cite{Caucal:2019uvr,Blanco:2020uzy,Karpenko:2024fgg}.

While the full linearized Boltzmann equation, \cref{eq:linBoltzmann}, obeys detailed balance and thus drives $\delta f$ toward thermal equilibrium, its high-energy limit, with \cref{eq:dC12_highenergy,eq:Broadening_highenergy}, lacks such an equilibration mechanism.
Radiation continuously degrades the energy of the initial perturbation, while momentum broadening keeps increasing transverse momentum. Energy-momentum conservation enforces momentum isotropization, but not thermalization. As a result, neither the high-energy limit of EKT nor the traditional jet-quenching evolution thermalizes the jet perturbation or the surrounding medium. On the other hand, the full linearized Boltzmann equation \cref{eq:linBoltzmann}, describes the full equilibration~\cite{Schlichting:2020lef,Mehtar-Tani:2022zwf}.

\textbf{The EKT parton shower} ---
Returning to the full evolution equation in \cref{eq:PartonShower_inhomo}, the implementation proceeds analogously to DGLAP evolution and to the high-energy limit above, but with a broader set of collision processes. For simplicity, we focus solely on gluons.

Real elastic collisions involve 
\begin{equation}
    \begin{split}
         \delta C^{2\leftrightarrow2,{\rm r}} &= \tfrac{1}{4p\nu}\int\rmd\Omega^{2\leftrightarrow2}|\mathcal M|^2\\ 
         &\times \Big[\delta f(t',\p') (n_{k'}\bar n_p\bar n_k - n_pn_k\bar n_{k'}) \\
         & \quad + \delta f(t',\k') (n_{p'}\bar n_p\bar n_k - n_pn_k\bar n_{p'}) \\
         & \quad - \delta f(t',\k) (n_{p}\bar n_{p'}\bar n_{k'} - n_{p'}n_{k'}\bar n_{p})\Big]\,,
    \end{split}
    \label{eq:LBT_22}
\end{equation}
and the corresponding no-scattering factor is
\begin{align} \label{eq:RealScatt_22}
        \Delta^{2\leftrightarrow2}(t,\p) & = \exp\left[-\int_{t_0}^t \rmd t' \delta C^{2\leftrightarrow2,{\rm v}}\right]\,,    \\
        \delta C^{2\leftrightarrow2,{\rm v}} & = \tfrac{1}{4p\nu}\int\rmd\Omega^{2\leftrightarrow2}|\mathcal M|^2 (n_k\bar n_{p'}\bar n_{k'}-n_{p'}n_{k'}\bar n_k)\,. \nonumber
\end{align}
The IR pole of the matrix element has to be regulated, for which we use Debye-like screening, $|t^2|,|u^2|>t^2_{\min}$, which affects the very-collinear region. Each $2\leftrightarrow2$ scattering creates three new particles corresponding to the three lines in \cref{eq:LBT_22}, one of which is a hole, i.e., a particle with negative weight.

The $1\leftrightarrow2$ collisions encompass a variety of splitting and merging processes~\cite{Soudi:2025lei}. The Sudakov factor is
\begin{equation}
    \begin{split}
        \ln\Delta^{1\leftrightarrow2}(t,p) &= -\int_{t_0}^t \rmd t'\,\rmd z\,(\Gamma_1(z,p)+\Gamma_2(z,p))\,,
    \end{split}
\end{equation}
with two real emission processes: the splitting $g(p)\to g(zp)+g(\bar zp)$, with rate (where $\bar z\!=\!1-z$):
\begin{equation}
  \Gamma_1(z,p)=\tfrac{1}{2}\Gamma(z,p)(1+n(zp)+n(\bar zp))\,,
\end{equation}
and the splitting/merging process $g(p)+g(\frac{zp}{\bar z})\leftrightarrow g(\frac{p}{\bar z})$:
\begin{equation}
  \Gamma_2(z,p)=\tfrac{1}{\bar z^3}\Gamma(z,\tfrac{p}{\bar z})(n(\tfrac{zp}{\bar z})-n(\tfrac{p}{\bar z}))\,.
\end{equation}
Merging arises from the negative terms in $\delta C^{1\leftrightarrow2,{\rm r}}$ and is handled analogously to holes. The IR poles of the splitting rates are regulated by requiring all particles to have energy greater than $E_{\min}$.

The parton-shower algorithm works as follows: the next scattering time is sampled from the combined Sudakov factor $\Delta(t_{\rm next},\p_0)$. Once a collision occurs ($t_{\rm next}<L$), the type of collision is determined according to the relative probabilities. Then the momenta of the new particles are drawn from the corresponding $\delta C^{\rm r}$; inelastic collisions produce two new particles, while elastic ones produce three, with loss contributions entering as particles of negative weight. Energy and momentum are conserved at each vertex. This procedure is iterated for all new particles until $t_{\rm next}>L$. A final Sudakov factor is applied to account for the absence of interactions between $t_{\rm last}$ and $L$. The rescattering of recoils and holes, together with merging, is necessary to reach equilibrium: the hole of a hole carries a positive weight, while the recoil of a hole is also negative. Similar combinations occur in mergers and in the combination of scatterings and merging. Following \cref{eq:PartonShower_inhomo}, we follow the trajectory of each particle by free-streaming them between collisions. 

\begin{figure}[h]
    \centering
    \includegraphics[width=\linewidth,page=1]{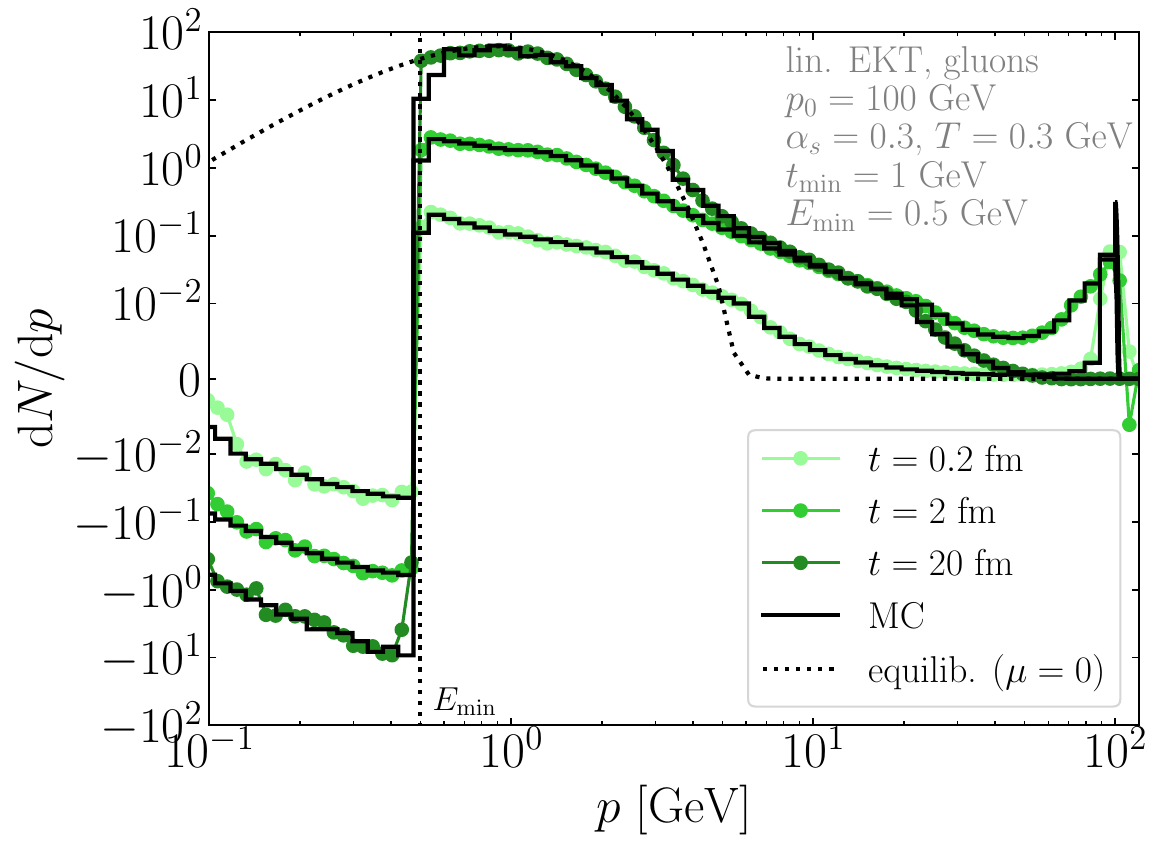}
    \caption{The gluon distribution as a function of the energy for different times. We establish equivalence between the traditional EKT solver and the new parton-shower implementation.}
    \label{fig:FullShower}
\end{figure} 

We implemented the corresponding parton-shower algorithm. The single-particle distribution is obtained by histogramming the shower particles,
\begin{equation}\label{eq:Histogram}
    \frac{\rmd N}{\rmd \p\rmd\x} = \frac{1}{N_{\rm ev}}\sum_{n=1}^{N_{\rm ev}}\sum_{i\in {\rm ev}_n}\frac{w_i}{\rmd \p\rmd\x}\,,
\end{equation}
where $N_{\rm ev}$ is the number of generated events. For simplicity, we focus on gluons, neglect quantum statistics, and use limits of AMY collisions (see App.~\ref{sec:Collision_kernels}). 

\Cref{fig:FullShower} shows the distribution of gluons as a function of the energy for different propagation times.
We chose typical parameters for jet quenching phenomenology, and a narrow 100 GeV single gluon perturbation.
At early times, the initial perturbation is peaked at $p_0=100$ GeV and vanishes over time as it fragments into softer particles.
The sharp cut at $E_{\min}$ originates from the IR regulation of $1\leftrightarrow2$.
We find numerical equivalence between the parton-shower (black lines) and the traditional differential-equation solver (points).
At late times, the perturbation approaches the perturbed equilibrium (dotted line),
\begin{equation}
    \begin{split}
        \delta f_{\rm eq}(\p) & = \left[\delta T\partial_T+\delta u^\nu\partial_{u^\nu}\right]n(p)\,.
    \end{split}
    \label{eq:deltaf_eq_mu0}
\end{equation}
Conservation laws $(p_0,0,0,p_0) = \int_{\p}p^\mu\,\delta f_{\rm eq}(\p)$ fix $\delta T$ and $\delta u^\mu$, and the equilibrium distribution is $n(p)\to e^{-p/T}$. This distribution changes sign at $\cos\theta_p^*=-1/3$.

\begin{figure}
    \centering
    \includegraphics[width=0.9\linewidth,page=1]{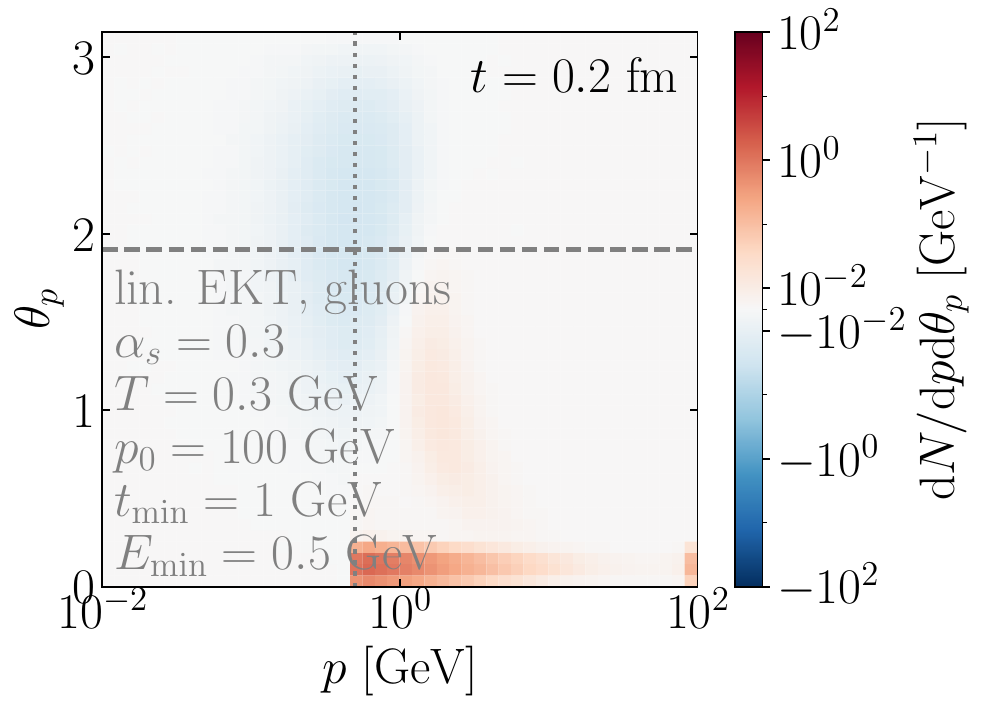}
    \includegraphics[width=0.9\linewidth,page=2]{figs/FullShower_2D.pdf}
    \includegraphics[width=0.9\linewidth,page=3]{figs/FullShower_2D.pdf}
    \includegraphics[width=0.9\linewidth,page=4]{figs/FullShower_2D.pdf}
    \caption{The gluon density in momentum space for different propagation times, where $\theta_p$ is relative to the perturbation.}
    \label{fig:FullShower_2D}
\end{figure}

\Cref{fig:FullShower_2D} shows the 2D number density in momentum space at different times.
At early times, the narrow perturbation peaks at $\theta_p=0$.
This peak first fragments into collinear particles and then broadens to wider angles. Later, the distribution approaches equilibrium see~\cref{eq:deltaf_eq_mu0}.

Number-conserving $2\leftrightarrow2$ scattering drives the perturbation toward an equilibrium with a finite chemical potential (\cref{eq:deltaf_eq_muNeq0}), whereas number-changing $1\leftrightarrow2$ processes lead to an equilibrium with vanishing chemical potential.
Interestingly, the out-of-equilibrium dynamics in $\theta_p$ initially follows $\mu\neq0$ because the early-stage angular broadening of soft fragments is dominated by elastic scatterings, before inelastic processes eventually drive the system to $\mu=0$ at later times. For more on $2\leftrightarrow2$ scatterings, see App.~\ref{sec:Elastic_scatterings}.

\Cref{fig:FullShower,fig:FullShower_2D} illustrate that the momentum distribution is not equilibrated in the phenomenologically relevant region, and thus the importance of keeping track of the full out-of-equilibrium evolution of jet perturbations. 

\begin{figure}
    \raggedright
    \includegraphics[width=\linewidth,page=3]{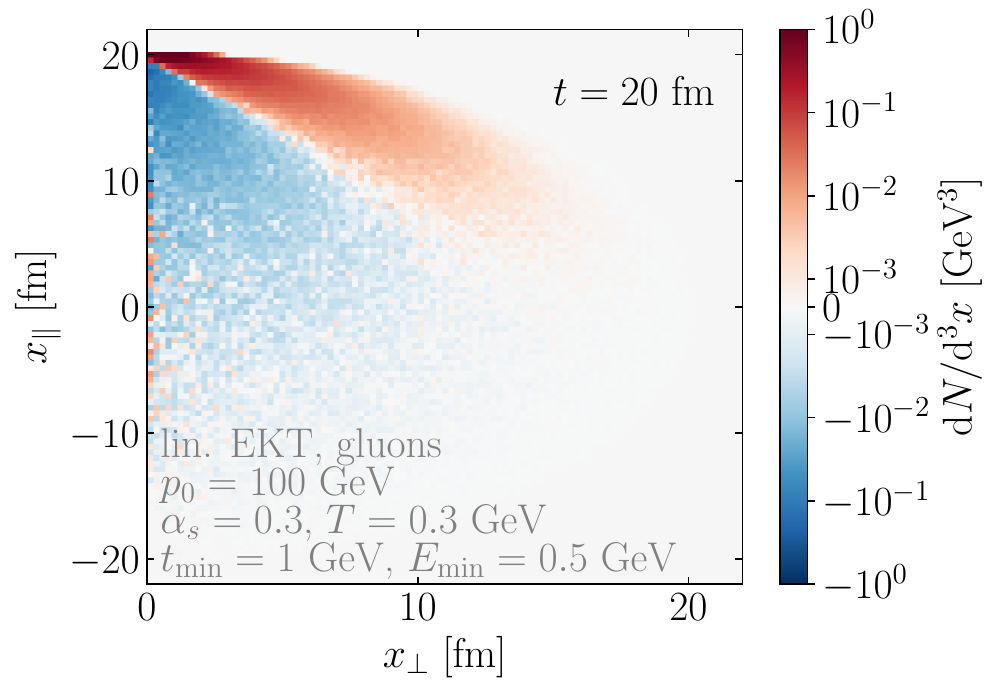}
    \caption{The evolution of the number density parallel and perpendicular to the initial perturbation.}
    \label{fig:FullShower_density}
\end{figure}

\textbf{Inhomogeneous perturbations.} \Cref{fig:FullShower_density} shows the spatial evolution of the gluon density at different time steps, with $x_\perp=\sqrt{x^2+y^2}$ and $x_\parallel=z$ for a perturbation oriented along the $z$-axis.
The perturbation propagates along $x_\parallel$ at the speed of light, and its fragments develop a non-trivial pattern of positive and negative density.
A Mach cone-like structure and a depletion trailing the perturbation are clearly visible, both characteristic of hydrodynamic responses~\cite{Casalderrey-Solana:2020rsj}.
We stress, however, that the evolution remains far from equilibrium, as is evident from \cref{fig:FullShower,fig:FullShower_2D}.

These results can serve as a benchmark for future implementations of inhomogeneities in traditional kinetic-theory solvers.
We also note that, although all particles in the simulation are massless and travel at the speed of light, collective modes are clearly excited, as seen in the plot.
This framework opens the way to future studies of how local equilibrium is approached in kinetic theories.

\textbf{Two-particle correlations} --- 
Parton showers also provide a model for estimating multi-particle correlations.
Although we have proved that we reproduce \cref{eq:linBoltzmann}, we have not shown, and it is generally unknown, how the multi-particle distributions themselves evolve in the EKT.
Instead, the parton shower assumes that correlations factorize in a Markovian picture, corresponding to the usual molecular chaos assumption of the Boltzmann equation.
Correlations are nonetheless not entirely absent: energy-momentum conservation at each vertex, together with $1\leftrightarrow2$ and $2\leftrightarrow2$ collisions, introduces non-trivial correlations.

\begin{figure}
    \centering
    \includegraphics[width=\linewidth,page=1]{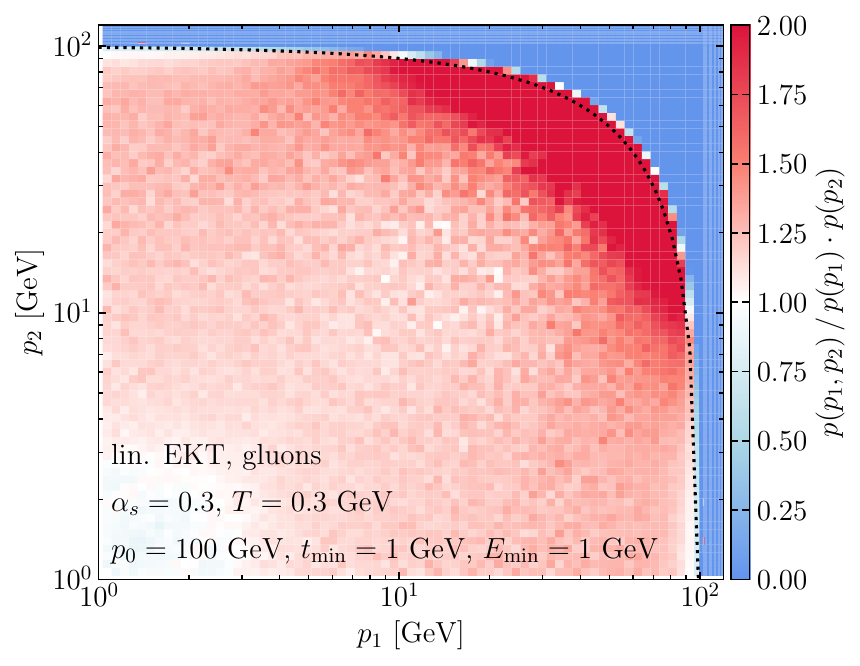}
    \includegraphics[width=\linewidth,page=1]{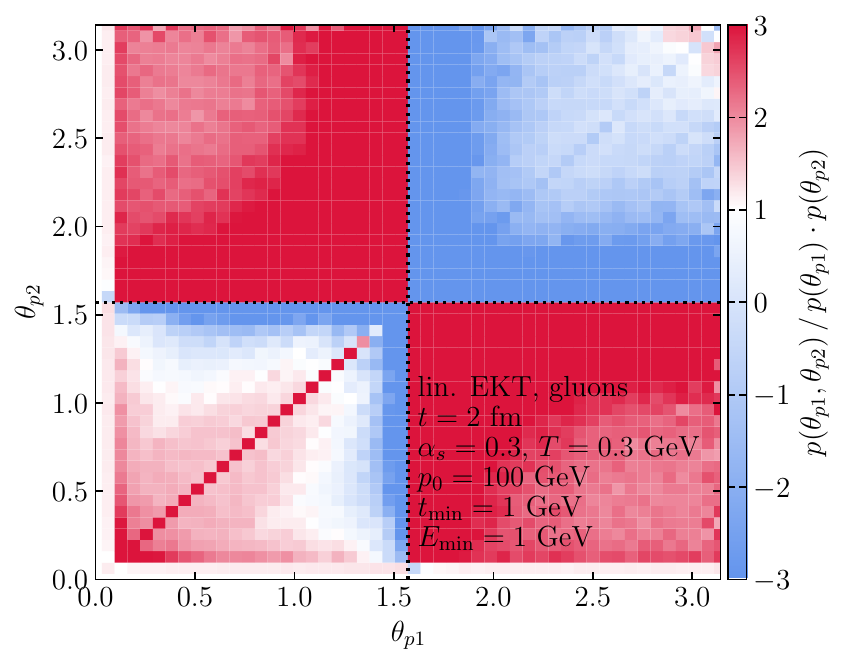}
    \caption{Two-particle distribution from the parton shower divided by the single-particle distributions at $t=2$ fm.}
    \label{fig:FullShower_corr}
\end{figure}

We evaluate the 2-particle distribution as
\begin{equation}\label{eq:Histogram_3D}
    \frac{\rmd N}{\rmd \p_1\rmd\x_1\rmd \p_2\rmd\x_2} = \frac{1}{N_{\rm ev}}\sum_{n=1}^{N_{\rm ev}}\sum_{i\neq j}^{ {\rm ev}_n}\frac{w_iw_j}{\rmd\p_1\rmd\x_1\rmd\p_2\rmd\x_2}\,.
\end{equation}
The upper panel of \cref{fig:FullShower_corr} shows the 2-particle distribution as a function of energy, normalized by the expectation of molecular chaos (single-particle) so that only genuine correlations remain.
Energy-momentum conservation, $p_1+p_2<p_0$, clearly bounds the accessible phase space (dotted line), and non-trivial correlations appear whenever $p_1,p_2\lesssim p_0$.
Interestingly, the correlations diminish for $p_1,p_2\approx T$ and for $p_1\ll p_2$ or $p_1\gg p_2$.

The lower panel of \cref{fig:FullShower_corr} shows the same $2$-particle distribution as a function of the angles relative to the jet axis.
The checkerboard pattern originates from isotropization: particles are more likely to be found moving in opposite directions relative to the jet (hence the boundary at $\pi/2$ shown with dotted lines), producing correlation in the red rectangle and anti-correlation in the blue.
Close to the jet axis, a non-trivial pattern emerges from $2\leftrightarrow2$ scatterings, whereas collinear fragments from $1\leftrightarrow2$ are in the diagonal.

\textbf{Conclusions} ---
We reformulated the linearized effective kinetic theory of QCD as a parton shower, which combines both $1\leftrightarrow2$ and $2\leftrightarrow2$ collisions, and demonstrated its equivalence to a numerical solution to the partial differential equation.
The high-energy limit of this parton shower reproduces the BDMPS-Z evolution equations for jet quenching, while the full parton shower describes the equilibration of a perturbation in a thermal background.
Our parton-shower implementation allows for a straightforward extension to inhomogeneous perturbations, which are difficult to study with traditional numerical methods.
Using this framework, we studied the equilibration of a minijet in a thermal background, showing for the first time how spatial inhomogeneities develop structures similar to Mach cones and a negative diffusion wake even when the system is far from equilibrium.
Because the QGP fireball created in heavy-ion collisions is a rapidly evolving, out-of-equilibrium system, the dynamics of jet-medium interactions cannot be captured by a simple superposition of decoupled jet quenching and hydrodynamic response models.
Our work establishes a unified framework to study the complete, real-time dynamics of jet-medium interactions, tracking the non-equilibrium evolution and ultimate equilibration of the medium response on a microscopic footing.
As the experimental precision of jet-substructure and correlation measurements at both the LHC and RHIC continues to advance~\cite{Cunqueiro:2021wls,CMS:2021otx,ATLAS:2024prm,CMS:2025dua,CMS:2026mur,STAR:2017hhs}, our framework will be crucial for interpreting these experimental signatures of the jet wake and extracting the transport properties of the QGP.

\begin{acknowledgements}
    We greatly appreciate discussions with Aleksas Mazeliauskas.
    The work of A.T. is supported by the Norwegian Research Council FRIPRO PreciseJets project.
    I.S. acknowledges support by the Deutsche Forschungsgemeinschaft (DFG, German Research Foundation) through the CRC-TR 211 `Strong-interaction matter under extreme conditions'--project number 315477589--TRR 211.
\end{acknowledgements}

\appendix

\section{Collision kernels}
\label{sec:Collision_kernels}
In this appendix, we briefly summarize the AMY collision kernels~\cite{Arnold:2002zm} (see also \cite{Zhou:2024ysb}), which are widely used to describe thermalization in QCD. The formulas in the main text regarding the parton shower rely only on the general structure of these kernels, not on the exact form of the matrix elements.

The elastic collision term is
\begin{equation}\label{eq:linC22}
    \begin{split}
        \delta C_a^{2\leftrightarrow2}(\p) & = \frac{1}{4|\bm p|\nu_a}\sum_{bcd}\int \rmd\Omega^{2\leftrightarrow2}|\mathcal M^{ab}_{cd}|^2 \delta \mathcal F^{ab}_{cd}\,,
    \end{split}
\end{equation}
where $\int \rmd\Omega^{2\leftrightarrow2}=\int_{k,p',k'}(2\pi)^4\delta^4(p^\mu+k^\mu-p'^\mu-k'^\mu)$, with $\int_p=\int_{\p}\frac{1}{2E_p}$, $\int_{\p}=\int\frac{\rmd^3\p}{(2\pi)^3}$, and the associated quantum-statistical factor is
\begin{equation}\label{eq:linStatF22}
    \begin{split}
        \delta\mathcal{F}^{ab}_{cd}(\p,\k;\p',\k') 
            & = \delta f_c(\p') \left[n_d\bar n_a\bar n_b \mp_c n_an_b\bar n_d\right]    \\
            & + \delta f_d(\k') \left[n_c\bar n_a\bar n_b \mp_d n_an_b\bar n_c\right]    \\
            & - \delta f_a(\p)  \left[n_b\bar n_c\bar n_d\mp_a n_cn_d\bar n_b\right]     \\
            & - \delta f_b(\k)  \left[n_a\bar n_c\bar n_d\mp_b n_cn_d\bar n_a\right] \,.
    \end{split}
\end{equation}
We omit the momentum dependence, which can be inferred from the indices $(a,\p)$, $(b,\k)$, $(c,\p')$, $(d,\k')$; $\bar n=1\pm n$; the sign $\pm_a$ varies for bosons and fermions; and the $\x$ dependence is local and implicit. The equilibrium background for bosons and fermions is $n_a(t,\x,\p)=1/(e^{p/T}\mp1)$. $|\mathcal M^{ab}_{cd}|^2$ are the HTL-regulated leading-order $2\leftrightarrow2$ matrix elements; in our numerics we use the LO QCD $gg\to gg$ element $|\mathcal M^{gg}_{gg}|^2=16g_s^4d_AN_c^2(3-\frac{us}{t^2}-\frac{ut}{s^2}-\frac{tu}{s^2})$ with Debye screening $t_{\min}\propto m_D^2$~\cite{Zapp:2008gi,Auvinen:2009qm,Li:2010ts,He:2015pra}.

Inelastic collisions are
\begin{equation}\label{eq:linC12}
    \begin{split}
        \delta C_a^{1\leftrightarrow2} 
            & = \sum_{bc}\int^1_0\rmd z\Big[\tfrac{1}{z^3}\Gamma^c_{ab}\left(z,\tfrac{\bm p}{z}\right) \delta \mathcal F^c_{ab}\left(\tfrac{\p}{z};\p,\tfrac{\bar z\p}{z}\right) \\
            & - \tfrac12\Gamma^a_{bc}(z,\bm p)\delta\mathcal F^a_{bc}(\p;z\p,\bar z\p)\Big] \,,
    \end{split}
\end{equation}
where the corresponding statistical factor is
\begin{equation}\label{eq:linStatF12}
    \begin{split}
        \delta\mathcal F^a_{bc}(\p;\k,\p') 
            & = \delta f_a\left[\bar n_b\bar n_c \mp_a n_bn_c\right]    \\
            & - \delta f_b\left[n_c\bar n_a \mp_b n_a\bar n_c\right]    \\
            & - \delta f_c\left[n_b\bar n_a \mp_c n_a\bar n_b\right]\,, \\
    \end{split}
\end{equation}
with $(a,\p),(b,\k),(c,\p')$. $\Gamma^a_{bc}$ are the AMY splitting rates. For simplicity, our numerics use the high-energy limit $\Gamma(z,p)=\frac{\alpha_s}{2\pi}\frac{C_A}{z\bar z}\sqrt{\frac{C_A\hat q}{z\bar z p}}$, with $\hat q=2\alpha_sTm_D^2$ and $m_D^2=4\pi\alpha_sT^2$, matching the BDMPS-Z result in the soft limit and the $\hat q$-approximation limits~\cite{Arnold:2008iy,Arnold:2009mr}.

\section{Inhomogeneous perturbations}
\label{sec:Inhomogeneous_perturbations}

In this section, we show how to include space-time inhomogeneities (the gradient term in \cref{eq:linBoltzmann}) in the parton shower. This allows us to study the evolution of spatial perturbations within the parton-shower framework, necessary to describe hydrodynamization.

For collisionless systems, $\delta C\to0$, \cref{eq:linBoltzmann} reduces to the free propagation of the initial condition, $\delta f^{(0)}(t,\x,\p) = \delta f(t_0,\x-\hat{\p}(t-t_0),\p)$. We thus introduce comoving coordinates that follow $\p$ from $t_0$ to $t$: $\x^{(p)}$,
\begin{equation}
    \begin{split}
        \x(t,\x^{(p)},\p) & =\x^{(p)}+\hat{\p}(t-t_0)\,,      \\
        \delta f(t,\x,\p) & =\delta\tilde f(t,\x^{(p)},\p)\,.
    \end{split}
    \label{eq:comving1}
\end{equation}
In these coordinates, the gradient term is absent\footnote{
    We thank Aleksas Mazeliauskas for pointing out this trick.},
\begin{equation}\label{eq:LinBoltzmannCoM}
    \partial_t\delta\tilde f(t,\x^{(p)},\p)=\delta\tilde C[\{n,\delta f\}](t,\x^{(p)},\p)\,.
\end{equation}
The collision term in comoving coordinates is
\begin{equation}
    \begin{split}
         & \delta\tilde C(t,\x^{(p)},\p) = \delta C[\delta f(t,\x=\x^{(p)}+\hat{\p}(t-t_0),\k)]    \\
         & \quad = \delta C[\delta \tilde f(t,\x^{(k)}=\x^{(p)}+(\hat{\p}-\hat{\k})(t-t_0),\k)]\,,
    \end{split}
\end{equation}
and it involves integrals over $\delta f$ with various momentum arguments, which we denote generically by $\k$. Different momentum arguments correspond to different comoving coordinates. Finally, we expressed the collision term with the comoving distribution.

In comoving coordinates, \cref{eq:LinBoltzmannCoM} is similar to the spatially homogeneous case, and the parton shower is known
\begin{equation}
    \begin{split}
         & \delta\tilde f(t,\x^{(p)},\p) = \Delta(t-t_0,\p) \delta\tilde f(t_0,\x^{(p)},\p)                        \\
         & \quad+\int^t_{t_0}\rmd t'\Delta(t-t',\p)\, \delta C^{\rm r}[\{n,\delta\tilde f(t',\x^{(k,t')},\k)\}]\,,
    \end{split}
\end{equation}
where $\x^{(k,t')}=\x^{(p,t')}+(\hat{\p}-\hat{\k})(t'-t_0)$.
The no-collision factor $\Delta(t,\p)=\exp[-\int^t_{t_0}\rmd t'\delta C^{\rm v}[n](\p)]$ is independent of inhomogeneity. In the first line, the comoving coordinate is evaluated at $t$, while in the second line it is evaluated at $t'$. Finally, returning to lab coordinates,
\begin{align} \label{eq:LinBoltzmann_PS_inhomo}
     & \delta f(t,\x,\p) = \Delta(t-t_0,\p) \delta f(t_0,\x-\hat{\p}(t-t_0),\p)\\
     & \quad +\int^t_{t_0}\rmd t'\Delta(t-t',\p)\delta C^{\rm r}[\{n,\delta f(t',\x-\hat{\p}(t-t'),\k)\}]\,. \nonumber
\end{align}
\Cref{eq:LinBoltzmann_PS_inhomo} is equivalent to \cref{eq:linBoltzmann}, with the difference confined to the spatial argument of the collision kernel.

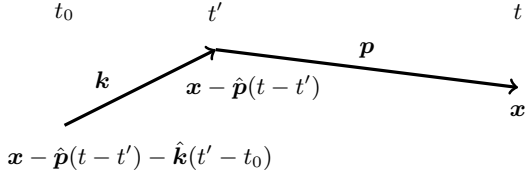
\begin{figure}[h]
    \centering
    \begin{tikzpicture}
        \draw (-3,-1.4) node {$\x-\hat\p(t-t')-\hat \k(t'-t_0)$} (-1.5,-0.5) node {$\x-\hat\p(t-t')$} (2,-0.8) node {$\x$};
        \draw (-3.5,-0.4) node {$\k$};
        \draw (0,0) node {$\p$};
        \draw [very thick,->] (-4,-1) -- (-2,0);
        \draw [very thick,->] (-2,0) -- (2,-0.5);
        \draw (-4,0.5) node {$t_0$} (-2,0.5) node {$t'$} (2,0.5) node {$t$};
    \end{tikzpicture}
    \caption{The spatial dependence of a collision using $(t,\x,\p)$.}\label{fig:CollisionKernelSpacetime}
\end{figure}

To interpret \cref{eq:LinBoltzmann_PS_inhomo}, we evaluate it with a delta-like initial condition, $\delta f(t_0,\x,\p)=\delta^3(\x-\x_0)\delta^3(\p-\p_0)$,
\begin{equation}
    \begin{split}\nonumber
        \delta f(t,\x,\p) & = \Delta(t-t_0,\p)\delta^3(\x-\hat\p(t-t_0)-\x_0)\delta^3_{pp_0}                   \\
                          & + \int_{t_0}^t\rmd t'\Delta(t-t',\p)\,\delta C^{\rm r}[\{n,\delta f^{(0)}\}]+\dots \\
        \delta f^{(0)}    & = \Delta(t'-t_0,\p_0)\delta^3_{kp_0}                                               \\
                          & \times \delta^3(\x-\hat\p(t-t')-\hat\k(t'-t_0)-\x_0)\,.
    \end{split}
\end{equation}
The first line describes the free streaming of the initial particle without collisions, ensured by the Sudakov factor. The second line represents a collision at $t'$, which must occur such that the original parton from $(t_0,\x_0,\p_0)$ ends up at $(t,\x,\p)$ given free propagation before and after the collision (see \cref{fig:CollisionKernelSpacetime} for an illustration).

\section{Parton shower with elastic collisions}
\label{sec:Elastic_scatterings}

\begin{figure}
    \centering
    \includegraphics[width=\linewidth]{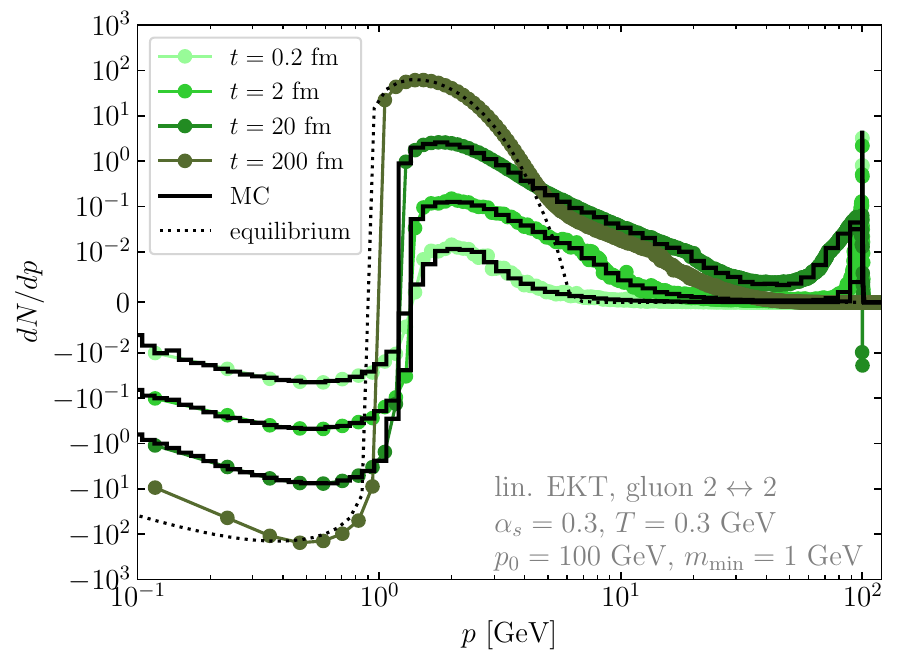}
    \includegraphics[width=\linewidth]{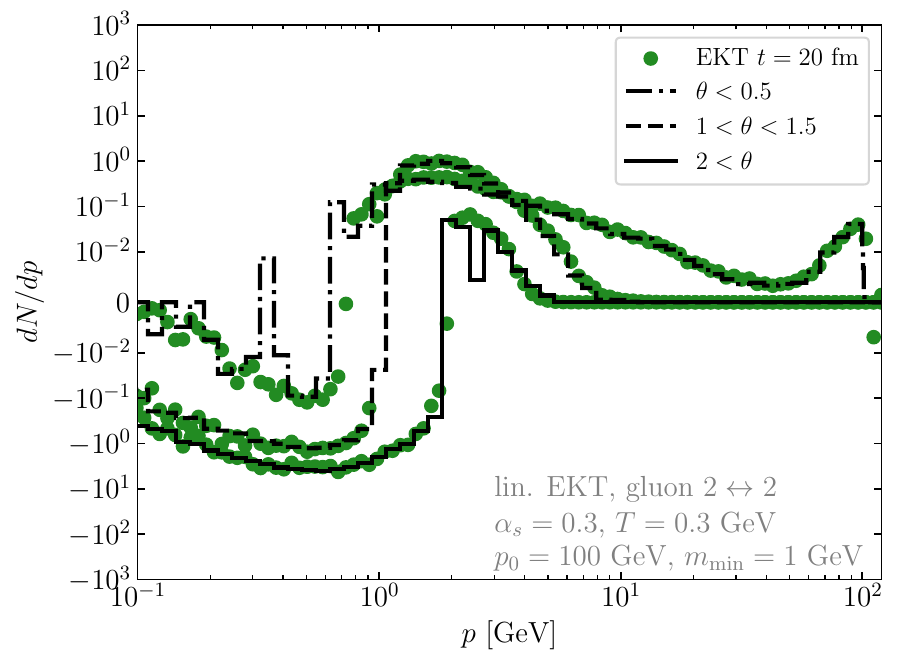}
    \caption{Energy distribution with only $2\leftrightarrow2$ collisions.}
    \label{fig:FullShower_22}
\end{figure}

In this section, we provide a direct comparison of $2\leftrightarrow2$ scatterings between the new parton shower and the traditional EKT solver, establishing numerical equivalence. We also show that the equilibrium and its approach change drastically with and without $1\leftrightarrow2$ scatterings.

\Cref{fig:FullShower_22} is analogous to \cref{fig:FullShower} but contains only $2\leftrightarrow2$ collisions. The initial energy peak broadens and decreases as it scatters into softer modes in the plasma. While 20 fm is enough to partially thermalize the same perturbation when $1\leftrightarrow2$ collisions are included (see \cref{fig:FullShower}), thermalization through $2\leftrightarrow2$ scatterings alone is much slower. The parton shower and the traditional differential equation solver yield identical distributions, since we employed the same matrix elements and phase-space cuts. It is important to note that the equilibrium (dotted line) is special as $2\leftrightarrow2$ collisions conserve particle number, introducing chemical potential,
\begin{equation}
    \begin{split}
        \delta f_{\rm eq}(\p) & = \left[\delta T\partial_T+\delta u^\nu\partial_{u^\nu}+\delta\mu\partial_{\mu}\right]n(p)\,.
    \end{split}
    \label{eq:deltaf_eq_muNeq0}
\end{equation}
Here, conservation laws $(p_0,0,0,p_0) = \int_{\p}p^\mu\,\delta f_{\rm eq}(\p)$ and $1 = \int_{\p}\delta f_{\rm eq}(\p)$, fix $\delta T,\delta u$, and $\delta\mu$. This equilibrium has a non-trivial boundary between positive and negative contribution $\cos\theta_p^*=-4(pp_0 - 3(p_0+p)T + 12T^2)/(3pp_0)$.

\bibliography{ref_vac,ref_med}

\end{document}